\documentclass[conference]{IEEEtran}
\IEEEoverridecommandlockouts
\pdfoutput=1
\usepackage{xcolor}
\usepackage{amsmath}

\usepackage{tikz}
\usepackage{pgfplots}
\usepackage{pgfplotstable}
\usepgfplotslibrary{fillbetween}
\usepgfplotslibrary{groupplots}
\pgfplotsset{compat=1.17}
\usetikzlibrary{arrows.meta}
\usepackage{subcaption}
\usepackage[font=footnotesize]{caption}
\usepackage{lipsum}

\usepackage[range-phrase=--,per-mode=symbol-or-fraction,range-units=single,list-units=single,detect-all]{siunitx}
\usepackage{csquotes}
\usepackage[hyphens]{url}
\usepackage{hyperref}
\usepackage[hyphenbreaks]{breakurl}
\usepackage[capitalise,noabbrev]{cleveref}
\Crefformat{figure}{#2Fig.~#1#3}
\Crefformat{equation}{#2Eq.~(#1)#3}

\ifCLASSINFOpdf
\else
\fi

\usepackage{graphicx}

\usepackage{acro}
\usepackage{mfirstuc}
\DeclareAcronym{MC}{short=MC, long=Molecular Communication}
\DeclareAcronym{FSI}{short=FSI, long=Fluid-Solid Interaction}
\DeclareAcronym{PS}{short=PS, long=Peak Systole}
\DeclareAcronym{ED}{short=ED, long=Early Diastole}
\DeclareAcronym{LD}{short=LD, long=Late Diastole}
\DeclareAcronym{CIR}{short=CIR, long=Channel Impulse Response}
\DeclareAcronym{TX}{short=TX, long=Transmitter}
\DeclareAcronym{RX}{short=RX, long=Receiver}
\DeclareAcronym{LES}{short=LES, long=Large Eddy Simulation}
\DeclareAcronym{DNS}{short=DNS, long=Direct Numerical Simulation}
\DeclareAcronym{RAS}{short=RAS, long=Reynolds Averaged Simulation}
\DeclareAcronym{3D}{short=3D, long=Three-Dimensional}
\DeclareAcronym{IoBNT}{short=IoBNT, long=Internet of Bio-Nano Things}
\DeclareAcronym{SPION}{short=SPION, long=SuperParamagnetic Iron-Oxide Nanoparticle}
\DeclareAcronym{MPPICFoam}{short=MPPIC, long=Multi-Phase Particle-In-Cell Foam}
\DeclareAcronym{CVS}{short=CVS, long=Cardiovascular System}
\DeclareAcronym{BNM}{short=BNM, long=Bio-Nano Machine}
\DeclareAcronym{SOI}{short=SOI, long=Start of Injection}
\DeclareAcronym{CFD}{short=CFD, long=Computational Fluid Dynamics}

\usepackage{titlesec}
\titlespacing\section{0pt}{3pt plus 2pt minus 2pt}{2pt plus 1pt minus 1pt}
\titlespacing\subsection{0pt}{3pt plus 2pt minus 2pt}{2pt plus 1pt minus 1pt}
\begin{document}

\setlength{\abovedisplayskip}{5pt}
\setlength{\belowdisplayskip}{5pt}

%\title{Advanced Plaque Modeling for Atherosclerosis Detection using Molecular Communication Channels}
\title{Advanced Plaque Modeling for Atherosclerosis Detection Using Molecular Communication\
\thanks{This work was supported by the German Research Foundation (DFG) as part of Germany's Excellence Strategy—EXC 2050/1—Cluster of Excellence "Centre for Tactile Internet with Human-in-the-Loop" (CeTI) of Technische Universität Dresden under project ID 390696704 and the Federal Ministry of Education and Research (BMBF) in the programme of "Souverän. Digital. Vernetzt." Joint project 6G-life, grant numbers 16KISK001K and 16KISK002. 
This work was also partly supported by the project IoBNT, funded by the German Federal Ministry of Education and Research (BMBF) under grant number 16KIS1994.}
\vspace{-0.3cm}
}
\author{
\IEEEauthorblockN{Alexander Wietfeld\IEEEauthorrefmark{1}, Pit Hofmann\IEEEauthorrefmark{2}, Jonas Fuchtmann\IEEEauthorrefmark{4}, Pengjie Zhou\IEEEauthorrefmark{2}, Ruifeng Zheng\IEEEauthorrefmark{2},\\ Juan A. Cabrera\IEEEauthorrefmark{2}, Frank H.P. Fitzek\IEEEauthorrefmark{2}\IEEEauthorrefmark{3}, and Wolfgang Kellerer\IEEEauthorrefmark{1}}
\IEEEauthorblockA{
\IEEEauthorrefmark{1}Chair of Communication Networks, Technical University of Munich, Germany \\
\IEEEauthorrefmark{2}Deutsche Telekom Chair of Communication Networks, Technische Universität Dresden, Germany \\
\IEEEauthorrefmark{3}Centre for Tactile Internet with Human-in-the-Loop (CeTI), Dresden, Germany\\
\IEEEauthorrefmark{4}MITI, Department of Clinical Medicine, TUM School of Medicine and Health, Technical University Munich, Germany\\
\texttt{\{alexander.wietfeld,jonas.fuchtmann,wolfgang.kellerer\}@tum.de,}\\ 
\texttt{\{pit.hofmann,pengjie.zhou,ruifeng.zheng,juan.cabrera,frank.fitzek\}@tu-dresden.de}
}
}

\markboth{Journal of \LaTeX\ Class Files,~Vol.~14, No.~8, August~2015}%
{Shell \MakeLowercase{\textit{et al.}}: Bare Demo of IEEEtran.cls for IEEE Journals}

\maketitle

\begin{abstract}
As one of the most prevalent diseases worldwide, plaque formation in human arteries, known as \textit{atherosclerosis}, is the focus of many research efforts. Previously, molecular communication (MC) models have been proposed to capture and analyze the natural processes inside the human body and to support the development of diagnosis and treatment methods. In the future, synthetic MC networks are envisioned to span the human body as part of the Internet of Bio-Nano Things (IoBNT), turning blood vessels into physical communication channels. By observing and characterizing changes in these channels, MC networks could play an active role in detecting diseases like \textit{atherosclerosis}. 
In this paper, building on previous preliminary work for simulating an MC scenario in a plaque-obstructed blood vessel, we evaluate different analytical models for non-Newtonian flow and derive associated channel impulse responses (CIRs). 
Additionally, we add the crucial factor of flow pulsatility to our simulation model and investigate the effect of the systole-diastole cycle on the received particles across the plaque channel. We observe a significant influence of the plaque on the channel in terms of the flow profile and CIR across different emission times in the cycle. These metrics could act as crucial indicators for early non-invasive plaque detection in advanced future MC methods.
\end{abstract}

\begin{IEEEkeywords}
Internet of Bio-Nano Things, Plaque Formations, Molecular Communication, OpenFOAM, Simulation
\end{IEEEkeywords}

\IEEEpeerreviewmaketitle

\section{Introduction}

% - our paper on plaque \cite{Hofmann2024}
% - our data set ~\cite{Hofmann2024dataet}

\IEEEPARstart{M}{olecular} communication is envisioned to play an important role in future communication networks as a biocompatible, energy-efficient way to transmit information in biological, micro- and nanoscale environments.
In a future \ac{IoBNT} so-called \acp{BNM}, enabled by recent advances in nanotechnology and bioengineering, could cooperate and communicate to achieve complex tasks such as supporting the diagnosis and treatment of diseases~\cite{Akyildiz2015internet}.
\textit{Atherosclerosis}, the formation of plaque inside of human arteries, is the underlying cause for about \SI{30}{\percent} of all deaths worldwide~\cite{preedyHandbookDiseaseBurdens2010}. In particular, plaques in the carotid artery have a prevalence of approximately \SI{21}{\percent} in the world's population between 30 and 79 years of age~\cite{song2020global}.
While there are several diagnosis and treatment approaches, they all require high-effort on-site procedures and are usually restricted to patients who are already symptomatic, as structured screening is rare~\cite{thaparDiagnosisManagementCarotid2013}.
Therefore, \textit{atherosclerosis} has become a topic among \ac{MC} researchers.
One approach seeks to utilize \textit{natural} \ac{MC} models to further our understanding of the initial stages of plaque growth, \textit{atherogenesis}. Felicetti \emph{et al.} have framed the process as a communication system, in which blood platelets act as \acp{TX} towards the endothelial cells at the blood vessel walls~\cite{felicettiModelingCD40BasedMolecular2014a}. Their work looks into detailed modeling, simulation, and lab experiments to analyze the breakdown of this natural process, which leads to plaque build-up, and to support the development of diagnosis and treatment methods.
In our previous work~\cite{Hofmann2024}, we proposed a second perspective on bringing together \ac{MC} research and \textit{atherosclerosis} based on \textit{synthetic} \ac{MC} networks that will be part of a future \ac{IoBNT} inside the human body. As \acp{BNM} will populate the human \ac{CVS} and emit and receive nanoparticles, the human body and its blood vessels become physical communication channels. As the human body changes, potentially in unhealthy ways, these communication channels may change in a way that we can detect in a joint communication and sensing approach. We simulated a simplified blood vessel scenario with a growing plaque using OpenFOAM and demonstrated that the obstacle significantly affects the communication channel.
In this work, we want to build upon the preliminary analysis in~\cite{Hofmann2024} in the following ways:
    %\begin{itemize}
        \textit{(i)} We will highlight possible analytical approaches to model different parts of the scenario, such as non-Newtonian fluid flow models, compare and contrast the resulting flow profiles, and discuss the limitations of analytical models in this complex scenario.
        \textit{(ii)} We derive analytical \acp{CIR} for the non-Newtonian flow models in the context of the proposed communication system structure and incorporate the Venturi effect in our analytical model.
        \textit{(iii)} We simulate the complex effects of a pulsatile inlet flow, carefully modeled after a realistic systole-diastole cycle of the heart, and apply it to our existing simulation. We demonstrate that our findings in~\cite{Hofmann2024} can be replicated for the pulsed flow and across different particle release times within the cycle.
    %\end{itemize}
    
This publication is connected to the release of our simulation data set, including the pulsatile inlet flow modeling~\cite{Hofmann2024dataet}.

\begin{figure}
    \centering
    \begin{tikzpicture}[scale=0.5, transform shape]
    \draw[fill=orange!20] (6.5,0) -- (7,1) -- (9,1) -- (9.5,0);
        \draw[thick,->,>={Triangle}] (0,1.5) -- (0,2.5);
        \draw[thick,->,>={Triangle}] (0,1.5) -- (1,1.5);
        \draw[thick] (0,1.5) circle (3pt);
        \node at (1,1.5+0.2) {x};
        \node at (0+0.3,2.5) {y,r};
        \node at (0-0.2,1.5-0.2) {z};
        \draw[blue] (1.5,0) -- (1.5,3);
        \begin{scope}[shift={(1.5,1.5)}]
        \draw[blue,domain=-1.5:1.5,samples=100] plot ({2.5*(1-(\x)^2/2.25)},\x);
        \draw[->,blue,>={Triangle}] (0,0) -- (2.5,0);
        \draw[->,blue,>={Triangle}] (0,-1) -- (1.39,-1.0);
        \draw[->,blue,>={Triangle}] (0,1) -- (1.39,1.0);
        \draw[->,blue,>={Triangle}] (0,-0.5) -- (2.22,-0.5);
        \draw[->,blue,>={Triangle}] (0,0.5) -- (2.22,0.5);
        \end{scope}
        \draw[black, fill=gray!30] (5, 1.5) ellipse [x radius=0.4, y radius=1.5];
        \draw[thick,fill=black] (5,1.5) circle (2pt);
        \draw[<->,>={Triangle[scale=0.5]}] (5,1.55) -- (5,3);
        \node at (5.15,2.25) {$r_c$};
        \draw[black, fill=gray!30] (11, 1.5) ellipse [x radius=0.4, y radius=1.5];
        \draw[thick] (0,3) -- (12,3);
        \draw[thick] (0,0) -- (12,0);
        \draw[dotted] (4,1.5) -- (12,1.5);
        \draw[thick,fill=black] (11,1.5) circle (2pt);
        \draw[<->,>={Triangle[scale=0.5]}] (5,3.25) -- (11,3.25);
        \node at (8,3.45) {$l_c$};
        \draw[dotted] (11,0) -- (11,3);
        
        \draw[thick,fill=green,green] (6.2,0.5) circle (1.5pt);
        \draw[thick,fill=green,green] (6.5,2.2) circle (1.5pt);
        \draw[thick,fill=green,green] (6.8,1.3) circle (1.5pt);
        %\draw[thick,fill=green,green] (7.1,0.8) circle (1.5pt);
        \draw[thick,fill=green,green] (7.4,1.9) circle (1.5pt);
        \draw[thick,fill=green,green] (7.7,2.7) circle (1.5pt);
        \draw[thick,fill=green,green] (8.3,2.5) circle (1.5pt);
        \draw[thick,fill=green,green] (8.7,1.6) circle (1.5pt);
        \node at (1,-1.5) {Pulsatile parabolic};
        \node at (1,-1.85) {flow profile};
        \draw[thin] (1.39+1.5,-1.0+1.5) -- (1.0,-1.2);
        \node at (7,-2.3) {Circular uniform transmitter};
        \draw[thin] (5,0.4) -- (5.5,-2.0);
        \node at (10,-1.5) {Circular observing receiver};
        \draw[thin] (11,0.4) -- (11.7,-1.2);
        \node at (2.5,-2.3) {Reflective boundary};
        \draw[thin] (3.5,0) -- (2.4,-2.0);
        \node at (5,4) {Released particles (SPIONs)};
        \draw[thin] (6.5,2.2) -- (5,3.8); 
        \node at (10,4) {Plaque formation};
        \draw[thin] (10,3.8) -- (8.8,0.8); 
        \draw[<->,>={Triangle[scale=0.5]}]  (6.5,-0.25) -- (9.5,-0.25);
        \node at (8,-0.5) {$l_{\mathrm{p,outer}}$};
        \draw[<->,>={Triangle[scale=0.5]}]  (7,1.25) -- (9,1.25);
        \node at (8,1.5) {$l_{\mathrm{p,inner}}$};

        \draw[thick,fill=orange!20] (14,1.5) circle (1.5cm);
        \draw[thick,fill=white] (14,2) circle (1cm);

        \draw[dotted] (12.3,1.5) -- (15.7,1.5);
        \draw[dotted] (14,3.2) -- (14,-0.2);

        \draw[dotted] (6.5,0.5) -- (9.5,0.5);
        \draw[dotted] (12.3,0.5) -- (15.7,0.5);

        \draw[<->,>={Triangle[scale=0.5]}]  (8,0) -- (8,0.5);
        \node at (8.3,0.25) {$r_p$};
        \draw[<->,>={Triangle[scale=0.5]}]  (14,0) -- (14,0.5);
        \node at (14.3,0.25) {$r_p$};
        \draw[<->,>={Triangle[scale=0.5]}]  (14,1.5) -- (14,3);
        \node at (14.3,2.25) {$r_c$};

        \draw[opacity=0.1, fill=yellow] (5,0) -- (5,3) -- (6.5,3) -- (6.5,0);
        \draw[opacity=0.1, fill=red] (7,0) -- (7,3) -- (6.5,3) -- (6.5,0);
        \draw[opacity=0.1, fill=green] (7,0) -- (7,3) -- (9,3) -- (9,0);
        \draw[opacity=0.1, fill=blue] (9.5,0) -- (9.5,3) -- (9,3) -- (9,0);
        \draw[opacity=0.1, fill=yellow] (9.5,0) -- (9.5,3) -- (11,3) -- (11,0);

        \node at (5.75,-0.9) {\textcolor{yellow}{\textbf{1}}};
        \node at (6.75,-0.9) {\textcolor{red}{\textbf{2}}};
        \node at (8.0,-0.9) {\textcolor{green}{\textbf{3}}};
        \node at (9.25,-0.9) {\textcolor{blue}{\textbf{4}}};
        \node at (10.25,-0.9) {\textcolor{yellow}{\textbf{1}}};

        \node at (4,-0.9) {\textcolor{black}{\textbf{Regions:}}};
    \end{tikzpicture}
    \caption{Schematic of the considered \textit{atherosclerosis} scenario; adapted from~\cite{Hofmann2024}.}
    \label{fig:plaque_schematic}
    \vspace{-0.6cm}
\end{figure}
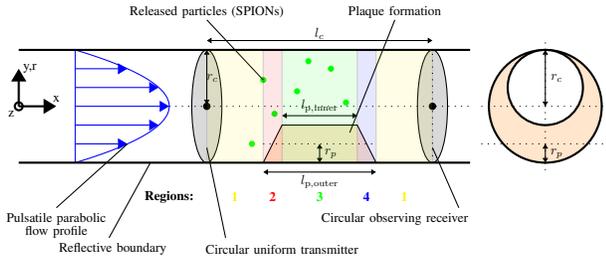

\section{MC Model Setup}
\label{sec:system_model}

Our \ac{MC} system model considers a 3D scenario, addressing the formation of plaque in human carotids, following our previous work in~\cite{Hofmann2024}, and schematically shown in Fig.~\ref{fig:plaque_schematic}.
The \ac{TX} is assumed to be circular. 
Upon release, the particles are uniformly distributed over the cross-section of the circular \ac{TX}.
After the propagation in the channel, the passive \ac{RX}, spanning the whole cross-section of the channel, counts the number of particles passing its cross-section.
The length of the channel $l_c = \SI{50.0}{\milli\meter}$, i.e., the distance between the circular \ac{TX} and passive \ac{RX}, is assumed to be constant.
The radius of the channel is $r_c = \SI{3.0}{\milli\meter}$, mimicking a human carotid~\cite{ferrara1995carotid}.
Following~\cite{Hofmann2024}, $r_p$ denotes the radial expansion of the plaque, related to the channel radius $r_c$. 
The default case assumes no plaque formation in the channel, i.e., no obstacle, so $r_p = 0 \times r_c$. We consider a constant inlet flow boundary condition, $u_{\rm avg} = \SI{34.2}{\centi\meter\per\second}$, and introduce a time-varying inlet flow, which will be described below. 
Additionally, we consider three different plaque sizes, $r_p = \{ 0.25,0.5,0.75 \} \times r_c$.
Finally, $l_{p,\rm outer} = \SI{20}{\milli\meter}$ and $l_{p,\rm inner} = \SI{10}{\milli\meter}$ denote the length of the plaque at the outside and the inside of the assumed human carotid, cf.~\cite[Fig.~1]{Hofmann2024} and Fig.~\ref{fig:plaque_schematic}.
%An overview of the simulation parameters can be found in Table~\ref{table:parameters}.

% \begin{table}[t]
%     \centering
%     \caption{Simulation parameters, following previous work in~\cite{Hofmann2024}.}
%     \begin{tabular}{|c|c|}
%     \hline 
%         \textbf{Parameter} & \textbf{Value} \\
%         \hline 
%         \hline
%         Channel length $l_c$ [\SI{}{\milli\meter}] & $50$ \\
%         \hline 
%         Radius $r_c$ [\SI{}{\milli\meter}] & $3.0$ \\
%         \hline 
%         Mean velocity $v_{\rm eff}$ [\SI{}{\centi\meter\per\second}] & $34.2$ \\
%         \hline 
%         Radial expansion plaque $r_p$ & $\{0,0.25,0.5,0.75\} \times r_c$ \\
%         \hline 
%         Length plaque inside $l_{p,\rm inner}$ [\SI{}{\milli\meter}] & $10$ \\
%         \hline 
%         Length plaque outside $l_{p,\rm outer}$ [\SI{}{\milli\meter}] & $20$ \\
%         \hline 
%     \end{tabular}
%     \label{table:parameters}
%     \vspace{-1.5em}
% \end{table}

%\subsection{Fluid Properties}

Our simulations consider human blood as the fluid phase, assuming a density of \SI{1050}{\kilo\gram\per\cubic\meter}~\cite{Kenner1989measurement}. 
Furthermore, we consider the Casson fluid model, as blood is a non-Newtonian fluid~\cite{Hofmann2024}.
%The following parameters are assumed for the Casson model, cf.~\cite{Casson_OpenFOAM}: $m=\SI{3.935e-6}{\meter\per\second}$, $\tau_0 = \SI{2.9032e-6}{\meter\per\square\second}$, $\nu_{max}=\SI{13.3333e-6}{\meter\per\second}$, and $\nu_{min}=\SI{3.9047e-6}{\meter\per\second}$.
The model parameters follow the used ones in~\cite{Casson_OpenFOAM}.
We assume laminar flow is present in all simulation scenarios. 
This preliminary assumption is intended for this initial study, as modeling turbulence for non-Newtonian fluids in a pipe with varying diameters and inlet velocity boundary conditions falls outside the scope of this paper.

%\subsection{Boundary Conditions}
%\label{subsec:boundary}

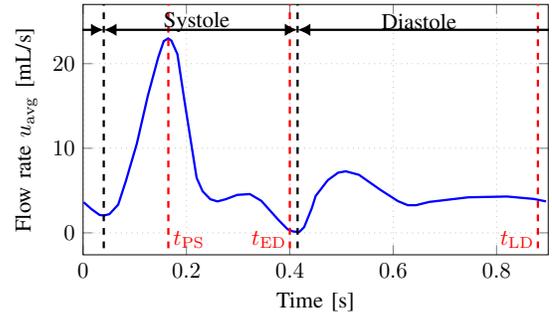
\begin{figure}
    \centering
    \begin{tikzpicture}[scale=0.85, every node/.style={transform shape}]
    
    \begin{axis}[
        xlabel={Time [\SI{}{\second}]},
        ylabel={Flow rate $u_{\rm avg}$ [\SI{}{\milli\liter\per\second}]},
        legend pos=south east,
        legend style={nodes={scale=0.8, transform shape}},
        width=\columnwidth,
        height = 5.5cm,
        grid style=dotted,
        grid=both,
        yticklabel style={/pgf/number format/fixed},
        xticklabel style={/pgf/number format/fixed}, 
        axis lines=box,
        xmin = 0.0,
        xmax = 0.9,
        ymax = 27,
    ]
    
    \addplot [mark=none, color=blue, line width = 1.0pt] table [x=x, y=y, col sep=comma] {figures/profile.csv};
    %\addlegendentry{$v_{\mathrm{inlet}}$}
    %\addlegendentry{Blood velocity profile shape}
\draw [red, dashed, line width=1pt] 
        (axis cs:0.165, -5) -- (axis cs:0.165, 30);
\node at (axis cs:0.205, -0.8) {\textcolor{red}{$t_{\mathrm{PS}}$}};
\draw [red, dashed, line width=1pt] 
        (axis cs:0.4, -5) -- (axis cs:0.4, 30);
\node at (axis cs:0.36, -0.8) {\textcolor{red}{$t_{\mathrm{ED}}$}};
\draw [red, dashed, line width=1pt] 
        (axis cs:0.88, -5) -- (axis cs:0.88, 30);
\node at (axis cs:0.84, -0.8) {\textcolor{red}{$t_{\mathrm{LD}}$}};
\draw [black, dashed, line width=1pt] 
        (axis cs:0.04, -5) -- (axis cs:0.04, 30);
\draw [black, dashed, line width=1pt] 
        (axis cs:0.415, -5) -- (axis cs:0.415, 30);
\draw [black, line width=1pt, {Latex[length=2mm, width=2mm]}-{Latex[length=2mm, width=2mm]}] 
        (axis cs:0.415, 24) -- (axis cs:0.04, 24);
\draw [black, line width=1pt, {Latex[length=2mm, width=2mm]}-{Latex[length=2mm, width=2mm]}] 
        (axis cs:0.415, 24) -- (axis cs:1, 24);
\draw [black, line width=1pt, {Latex[length=2mm, width=2mm]}-{Latex[length=2mm, width=2mm]}] 
        (axis cs:-0.415, 24) -- (axis cs:0.04, 24);
\node at (axis cs:0.65, 25) {Diastole};
\node at (axis cs:0.22, 25) {Systole};
    
    \end{axis}
    \end{tikzpicture}
    \vspace{-0.2cm}
    \caption{Human carotid artery pulsatile flow profile~\cite{Gay2008,Holdsworth1999}.}
    \label{fig:pulsatile}
    \vspace{-0.4cm}
\end{figure}

We apply the same boundary conditions as in our previous work in~\cite{Hofmann2024}, excluding the inlet velocity boundary condition. 
A detailed description of the remaining boundary conditions can be found in~\cite{Hofmann2024}.
We apply a human carotid pulsatile flow profile, cf.~Fig.~\ref{fig:pulsatile}, for the inlet velocity boundary condition using \textit{uniformFixedValue}, targeting future \ac{MC} applications in the \ac{IoBNT}. 
The human \ac{CVS} is influenced by the alternation between systole, the heart's active contraction phase, and diastole, the heart's passive relaxation phase. 
Both phases are considered, and the pulsatile flow is repeated every \SI{0.9}{\second}.

% \begin{figure}
%     \centering
%     \includegraphics[width=\linewidth]{figures/Model_1_b.pdf}
%     \vspace{-0.4cm}
%     \caption{Scheme of the considered \textit{atherosclerosis} scenario; adapted from~\cite{Hofmann2024}. \textcolor{red}{Can later show the different regions for the piecewise-defined radius as utilized in Eq. (\ref{eq:piecewise_radius}).}
%     }
%     \label{fig:plaque_schematic}
%     \vspace{-0.6cm}
% \end{figure}

\section{Analytical Approaches}
As a first step, we will investigate analytical modeling of the blood vessel and plaque obstruction scenario.
We focus on blood's non-Newtonian nature and compare the impact on the flow profile. Additionally, we will frame the non-Newtonian flow channel as an \ac{MC} channel and derive the associated \acp{CIR} at the \ac{RX}.

\subsection{Analytical Flow Profiles and Impulse Responses}

Blood is commonly modeled as a non-Newtonian fluid and will form velocity profiles that differ from the standard Poiseuille model. This is in part due to an effect called \textit{plug flow}, where the faster-moving red blood cells and other larger particles tend to accumulate in the center of the channel and flatten the front of the velocity profile~\cite{ashrafmansouriMathematicalModelingMicro2024}. This is generally relevant, but the effects are particularly important as the diameter of the blood vessel decreases, such as in the plaque-obstructed blood vessels we are concerned with in this work~\cite{wajihahReviewNonNewtonianFluid2023}.
There are numerous different viscosity models for non-Newtonian fluids, including the Casson and Bird-Carreau model, as used in~\cite{Hofmann2024}, for fluid dynamics simulations in a blood vessel. While these specific models are not suitable to derive closed-form analytical models for the flow profile or the \ac{CIR}, there exist other models incorporating very similar non-Newtonian effects, particularly the power-law and the Herschel-Bulkley models~\cite{guzelPredictingLaminarTurbulent2009}. For these models, also commonly used in blood rheology~\cite{wajihahReviewNonNewtonianFluid2023}, the Hagen-Poiseuille solution for a straight cylindrical pipe can be derived in closed form~\cite{guzelPredictingLaminarTurbulent2009}. In the following, we will introduce both models briefly, apply them to our \ac{MC} scenario, derive the \acp{CIR}, and subsequently compare them to the standard Newtonian model and to each other.

\subsubsection{Newtonian Fluid} Briefly, we will present the solutions for a Newtonian fluid. Using the fluid's shear stress $\tau$, consistency index $K$, and the shear rate $\dot{\gamma}$, the constitutive equation is given by~\cite{White1999fluid}
\begin{equation}
    \tau = K\cdot \dot{\gamma}.
\end{equation}
The resulting flow velocity profile over the channels radial coordinate $\rho = \sqrt{y^2 + z^2}$ is the well-known parabolic Poiseuille solution
\begin{equation}\label{eq:cir_newtonian}
    u(\rho) = u_0\left(1-\left(\frac{\rho}{r_c}\right)^2\right) ;\ 0\leq \rho \leq r_c.
\end{equation}
The \ac{CIR} $h(t)$ is defined as the fraction of particles that have reached the \ac{RX} at time $t$ for an instantaneous release of the particles at time $t=0$. The \ac{TX} is at $x=0$ and the \ac{RX} at $x=l_c$, extending to infinity along the rest of the channel. 

Before applying a specific model, we must determine which regime concerning flow and diffusion we are operating in. Following~\cite{jamali2019channel}, the so-called dispersion factor $\alpha$ is given by $\alpha = \frac{Dl_c}{u_{\rm avg} r_c^2}$, where $D$ denotes the diffusion coefficient of the particles. %, $l_c = \SI{50}{\milli\meter}$, and $r_c = \SI{3}{\milli\meter}$.
For the particles, we assume the commonly used \acp{SPION}~\cite{tietzeEfficientDrugdeliveryUsing2013} and their diffusion coefficient can be estimated from the average radius $r_\mathrm{SPION} = \SI{50}{\nano\meter}$~\cite{Bartunik2023channel}. The formula for the diffusion coefficient for particles suspended in a liquid~\cite{jamali2019channel} is $D = \frac{k_\mathrm{B}T}{6\pi\eta r_\mathrm{SPION}} = \SI{1.1e-12}{\square\meter\per\second}$,
where $k_\mathrm{B} = \SI{1.38e-23}{\joule\per\kelvin}$, $T=\SI{300}{\kelvin}$, and the dynamic viscosity of blood is set to an average value of $\eta = \SI{4e-3}{\pascal\second}$. In this case, the dispersion factor can be calculated to be $\alpha = 1.8\times10^{-8} \ll 1$, signifying the clear validity of the flow-dominated regime~\cite{jamali2019channel}.

Wicke \emph{et al.} derived the resulting \ac{CIR} for flow-dominated conditions~\cite{wickeExperimentalSystemMolecular2022}. Adapting their result for our scenario as shown in Fig.~\ref{fig:plaque_schematic} based on the system model in Section~\ref{sec:system_model} yields
\begin{equation}
    h(t) = 1-\frac{d}{u_0t};\ \mathrm{for}\ t\geq \frac{d}{u_0},
\end{equation}
where $u_0$ is the maximum flow speed in the middle of the channel and $u_0 = 2u_\mathrm{avg}$ in this case.
\subsubsection{Power-Law Fluid}\label{subsubsec:power-law}
The power-law non-Newtonian fluid model is defined by the constitutive equation
\begin{equation}
    \tau = K\cdot \dot{\gamma}^n.
\end{equation}
Here, $n$ is the power-law coefficient, and its inverse is $m=\frac{1}{n}$. The closed-form Hagen-Poiseuille solution for the flow velocity profile is given by~\cite{guzelPredictingLaminarTurbulent2009}
\begin{equation}\label{eq:cir_pl}
    u(\rho) = u_0\cdot\left(1-\left(\frac{\rho}{r_c}\right)^{m+1}\right);\ 0\leq \rho \leq r_c.
\end{equation}
Now, the goal is to derive a \ac{CIR} at the \ac{RX} similar to the Newtonian model. Given the initially uniform distribution of the particles in the \ac{TX} cross-section $f(\rho, \phi) = \frac{1}{\pi r_c^2}$, the initial concentration of particles can be expressed as $c_\mathrm{i}(\mathbf{x}) = f(\rho,\phi)\cdot \delta(x)$, where $\mathbf{x} = [\rho, \phi, x]$ and $\delta(\cdot)$ is the Dirac delta function. Following the \ac{CIR} derivation for the Newtonian case in~\cite{wickeExperimentalSystemMolecular2022}, we define an auxiliary variable $s = \frac{\rho^2}{r_c^2}$. As we are operating in the flow-dominated regime, the initial concentration is displaced over time according to the flow velocity profile $u(\rho)$ and then integrated over the \ac{RX} volume, assumed to be the infinite cylinder with $x>l_c$, $0\leq \phi < 2\pi$, $0\leq\rho\leq r_c$:
\begin{align}
    h(t) &= \iiint_V c_\mathrm{i}\left(\mathbf{x} - u(\rho)t\cdot \mathbf{\underline{e}}_z\right) \mathrm{d}V \nonumber\\
         &\!\!\!\!\!\!\overset{\mathrm{if}\ t\geq \frac{l_c}{u_0}}{=} \int_0^{r_c}\int_{l_c}^{u_0 t}\int_0^{2\pi} \frac{1}{\pi r_c^2} \delta(x-u(\rho)t)\rho \mathrm{d}\phi \mathrm{d}x \mathrm{d}\rho \nonumber\\
         &= \int_0^{r_c}\int_{l_c}^{u_0 t} \frac{2\rho}{r_c^2} \delta(x-u(\rho)t) \mathrm{d}x \mathrm{d}\rho \nonumber
\end{align}
\begin{equation}\label{eq:cir_deriv1}
         \overset{s=\frac{\rho^2}{r_c^2}}{=} \int_0^1\int_{l_c}^{u_0 t} \delta(x-u(\rho)t) \mathrm{d}x \mathrm{d}s.
\end{equation}
We can utilize the following property of the Dirac-delta function: $\delta(\varphi(s)) = \frac{1}{|\varphi^\prime (s_0)|}\delta(s-s_0)$, where $\varphi(s_0) = 0$. Then, $\varphi(s) = x - u_0t\left(1-s^{\frac{m+1}{2}}\right)$, $\varphi^\prime(s) = \frac{m+1}{2}u_0ts^{\frac{m-1}{2}}$ and
\begin{equation}
    \varphi(s_0) = 0\ \Leftrightarrow\  s_0 = \left(1-\frac{x}{u_0t}\right)^{\frac{2}{m+1}}.
\end{equation}
Since no particle travels with speed greater than $u_0$, we know that $0\leq x\leq u_0 t$ holds for our integration region and, therefore, $0\leq s_0\leq1$. We make use of the fact that $\int_0^1 \delta(s-s_0)ds = 1$ and can continue Eq. (\ref{eq:cir_deriv1}) as
\begin{multline}
    h(t)\! =\!\! \int_{l_c}^{u_0 t}\!\!\! \frac{1}{|\varphi^\prime(s_0)|} \mathrm{d}x\!=\!\! \int_{l_c}^{u_0 t}\!\!\! \frac{2}{(m+1)u_0t} \left(1-\frac{x}{u_0t}\right)^{\frac{1-m}{1+m}} \mathrm{d}x \\
    \!\!=\! \left[\!-\!\left(1-\frac{x}{u_0t}\right)^{\frac{2}{m+1}}\!\right]_{l_c}^{u_0t}\!\!\!\! =\! \left(1-\frac{l_c}{u_0t}\right)^{\frac{2}{m+1}}\!\!\!\!;\ \mathrm{for}\ t\!\geq\! \frac{l_c}{u_0}.
\end{multline}
For a different shape of the flow profile, the relationship between $u_0$ and $u_\mathrm{avg}$ also changes.
% \begin{equation}\label{eq:uavg}
%     u_\mathrm{avg} = \frac{1}{\pi r_c^2} \int_0^{r_c} \int_0^{2\pi} u(\rho) \rho \mathrm{d}\rho \mathrm{d}\phi,
% \end{equation}
For the power-law model, the result is~\cite{guzelPredictingLaminarTurbulent2009}
\begin{equation}
    \frac{u_0}{u_\mathrm{avg}} = \frac{m+3}{m+1}.
\end{equation}

\subsubsection{Herschel-Bulkley Model}

The Herschel-Bulkley model is more detailed than the power-law model, as it depends on one additional parameter. The constitutive equations are
\begin{align}
    \tau &= K\dot{\gamma}^n + \tau_\mathrm{y};\ &\mathrm{for}\ \tau > \tau_\mathrm{y}\\
    0 &= \dot{\gamma};\ &\mathrm{for}\ \tau\leq\tau_\mathrm{y},
\end{align}
with the yield stress $\tau_\mathrm{y}$. The Hagen-Poiseuille solution is given in~\cite{guzelPredictingLaminarTurbulent2009} as
\begin{equation}
    u(\rho) = 
    \begin{cases}
        u_0 & \rho < \rho_p\\
        u_0\cdot\left(1-\left(\frac{\rho-\rho_p}{r_c-\rho_p}\right)^{m+1}\right) & \rho_p\leq\rho\leq r_c.
    \end{cases}
\end{equation}
Here, $\rho_p = \zeta r_c$ is the plug radius that is calculated from the yield surface position $\zeta$, which is the zero of the so-called Buckingham polynomial 
% \begin{equation}
%     0 = \zeta^m - B(1-\zeta)^{m+1}\left[\frac{(1-\zeta)^2}{m+3} + \frac{2\zeta(1-\zeta)}{m+2} + \frac{\zeta^2}{m+1}\right],
% \end{equation}
on the interval $[0,1)$~\cite{guzelPredictingLaminarTurbulent2009}. 
% The Bingham number $B$ is given by $B=\left(\frac{\tau_\mathrm{y}}{K}\right)^m \frac{r_c}{u_\mathrm{avg}}$ and 
The relationship of $u_\mathrm{avg}$ to $u_0$ can be calculated as~\cite{guzelPredictingLaminarTurbulent2009}
\begin{equation}
    \frac{u_0}{u_\mathrm{avg}} = \frac{(m+2)(m+3)}{2\zeta^2 + 2(m+1)\zeta + (m+2)(m+1)}.
\end{equation}

For the sake of brevity, we now present the \ac{CIR} of the Herschel-Bulkley model, which we have derived in a very similar manner to the power-law model in Section~\ref{subsubsec:power-law}:
\begin{equation}\label{eq:cir_hb}
    h(t)\! =\! \left(\!\frac{\rho_p}{a} + \left(1-\frac{\rho_p}{a}\right)\!\left(1-\frac{d}{u_0 t}\right)^{\frac{1}{m+1}}\!\right)^2\!\!\!\!;\mathrm{for}\ t\geq \frac{d}{u_0t}. 
\end{equation}

\subsection{Venturi Effect}

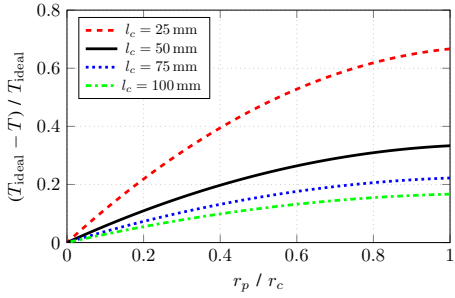
\begin{figure}
    \centering
    \begin{tikzpicture}[scale=0.7, every node/.style={transform shape}]
    
    \begin{axis}[
        xlabel={$r_p$ / $r_c$},
        ylabel={($T_{\mathrm{ideal}} - T$) / $T_{\mathrm{ideal}}$},
        legend pos=north west,
        legend style={nodes={scale=0.8, transform shape}},
        width=\columnwidth,
        height = 6cm,
        grid style=dotted,
        grid=both,
        yticklabel style={/pgf/number format/fixed},
        xticklabel style={/pgf/number format/fixed}, 
        axis lines=box,
        xmin = 0.0,
        xmax = 1.0,
        ymin = 0,
        ymax = 0.8,
    ]
    
    \addplot [mark=none, color=red, line width = 1.5pt, dashed] table [x expr=1- \thisrowno{0}, y expr=\thisrowno{1}, col sep=comma] {figures/venturi.csv};
    \addlegendentry{$l_c = \SI{25}{\milli\meter}$}

    \addplot [mark=none, color=black, line width = 1.5pt] table [x expr=1- \thisrowno{0}, y expr=\thisrowno{2}, col sep=comma] {figures/venturi.csv};
    \addlegendentry{$l_c = \SI{50}{\milli\meter}$}

    \addplot [mark=none, color=blue, line width = 1.5pt, dotted] table [x expr=1- \thisrowno{0}, y expr=\thisrowno{3}, col sep=comma] {figures/venturi.csv};
    \addlegendentry{$l_c = \SI{75}{\milli\meter}$}

    \addplot [mark=none, color=green, line width = 1.5pt, dash dot] table [x expr=1- \thisrowno{0}, y expr=\thisrowno{4}, col sep=comma] {figures/venturi.csv};
    \addlegendentry{$l_c = \SI{100}{\milli\meter}$}
    
    \end{axis}
    \end{tikzpicture}
    \vspace{-0.2cm}
    \caption{Relative reduction in traversal time due to the Venturi effect in the narrower channel with plaque of size $r_\mathrm{p}$.}
    \label{fig:venturi}
    \vspace{-0.6cm}
\end{figure}

While the fluid models above capture the impact of the non-Newtonian behavior of blood, we will also try to look at a very simple analytical approach to modeling the impact of the plaque. Modeling the flow profile's exact development across channel radius transitions analytically is extremely challenging. Therefore, we focus on one aspect, the \textit{Venturi effect}~\cite{schaschkeDictionaryChemicalEngineering2014}. The relationship between the volume flow $Q$, the cross-sectional area of the cylindrical channel $A=\pi r^2$, and the flow speed $u$ is given by $Q=Au$. Assuming constant volume flow, the change in channel radius along the $x$-axis, $r(x)$, due to the plaque will impact the flow speed according to $u(x) = u_0\frac{r_c^2}{r(x)^2}$. The radius can be expressed as a piecewise function
\begin{equation}\label{eq:piecewise_radius}
    r(x)\! =\! \begin{cases}
        r_c       & \mathrm{Region}\ 1\\
        r_c\! -\! r_\mathrm{p}\!\left(\frac{2x-(l_c-l_\mathrm{p,outer})}{l_\mathrm{p,outer} - l_\mathrm{p,inner}}\right)       &\mathrm{Region}\ 2\\
        r_\mathrm{p}        &\mathrm{Region}\ 3\\
        r_\mathrm{p}\! +\! (r_c\!-\!r_\mathrm{p})\!\left(\frac{2x-(l_c+l_\mathrm{p,outer})}{l_\mathrm{p,outer} - l_\mathrm{p,inner}}\right)      &\mathrm{Region}\ 4.
    \end{cases}
\end{equation}
The different regions along the channel are specified in Fig.~\ref{fig:plaque_schematic}. Assuming a very simplified model in which the maximum flow speed is increased according to the Venturi effect for $r(x)$, we can calculate the minimum time for a particle to traverse the channel
\begin{equation}\label{eq:venturi}
    T = \int_0^{l_c} \frac{\mathrm{d}x}{u(x)} = \frac{1}{u_0} \int_0^{l_c}\frac{r_c^2}{r(x)^2} \mathrm{d}x.
\end{equation}
We can see that since $\frac{r_c^2}{r(x)^2} \geq 1$, $T \leq T_\mathrm{ideal} = \frac{l_c}{u_0}$ will hold and the particles will arrive faster at the \ac{RX} in the presence of a plaque. In Fig.~\ref{fig:venturi}, we plot the relative reduction in traversal time over the relative plaque radius. It can be seen that the presence of a plaque can cause a significant speed increase that could represent an important viable metric for early plaque detection using an advanced \ac{IoBNT} system. Additionally, the results are shown for various channel lengths between \ac{TX} and \ac{RX}. This shows that the proximity to the plaque plays an important role in determining the magnitude of the change. A certain density of \acp{BNM} in the \ac{CVS} must be guaranteed to enable detection. However, this could also provide additional information about locating the plaque in relation to the \ac{TX} and \ac{RX}.

\subsection{Model Comparison and Flow Profile Simulation}

\begin{figure}
    %\includegraphics[height=6cm]{figures/flow_profile_comparison.pdf}
    %\hfill
    %\includegraphics[height=6cm]{figures/cir_comparison.pdf}
    \begin{subfigure}{0.49\columnwidth} 
    \centering
    \begin{tikzpicture}[scale=1.0, every node/.style={transform shape}]
    
    \begin{axis}[
        xlabel={$\rho \: [\SI{}{\meter}]$},
        ylabel={$u(\rho) \: [\SI{}{\meter\per\second}]$},
        legend pos=south east,
        legend style={nodes={scale=0.56, transform shape}},
        width=\columnwidth,
        height = 6cm,
        grid style=dotted,
        grid=both,
        yticklabel style={/pgf/number format/fixed},
        xticklabel style={/pgf/number format/fixed}, 
        axis lines=box,
        xmin = -3e-3,
        xmax = 3e-3,
        ymin = 0,
        ymax = 0.7,
    ]
    
    \addplot [mark=none, color=black, line width = 1.0pt] table [x expr=\thisrowno{0}, y expr=\thisrowno{1}, col sep=comma] {figures/flow_profiles.csv};
    \addlegendentry{Newtonian}

    \addplot [mark=none, color=red, line width = 1.0pt] table [x expr=\thisrowno{0}, y expr=\thisrowno{2}, col sep=comma] {figures/flow_profiles.csv};
    \addlegendentry{Power-law}

    \addplot [mark=none, color=blue, line width = 1.0pt, dashed] table [x expr=\thisrowno{0}, y expr=\thisrowno{3}, col sep=comma] {figures/flow_profiles.csv};
    \addlegendentry{Herschel-Bulkley}
    
    \end{axis}
    \end{tikzpicture}
    \end{subfigure}% 
    \begin{subfigure}{0.49\columnwidth} 
    \centering
    \begin{tikzpicture}[scale=1.0, every node/.style={transform shape}]
    
    \begin{axis}[
        xlabel={$t \: [\SI{}{\second}]$},
        ylabel={Fraction of received particles},
        legend pos=south east,
        legend style={nodes={scale=0.56, transform shape}},
        width=\columnwidth,
        height = 6cm,
        grid style=dotted,
        grid=both,
        yticklabel style={/pgf/number format/fixed},
        xticklabel style={/pgf/number format/fixed}, 
        axis lines=box,
        xmin = 0,
        xmax = 1.0,
        ymin = 0,
        ymax = 1.0,
    ]
    
    \addplot [mark=none, color=black, line width = 1.0pt] table [x expr=\thisrowno{0}, y expr=\thisrowno{1}, col sep=comma] {figures/analytical_cir.csv};
    \addlegendentry{Newtonian}

    \addplot [mark=none, color=red, line width = 1.0pt] table [x expr=\thisrowno{0}, y expr=\thisrowno{2}, col sep=comma] {figures/analytical_cir.csv};
    \addlegendentry{Power-law}

    \addplot [mark=none, color=blue, line width = 1.0pt, dashed] table [x expr=\thisrowno{0}, y expr=\thisrowno{3}, col sep=comma] {figures/analytical_cir.csv};
    \addlegendentry{Herschel-Bulkley}
    
    \end{axis}
    \end{tikzpicture}
    \end{subfigure}% 
    \vspace{-0.2cm}
    \caption{Comparison of three analytical fluid flow models, two incorporating non-Newtonian effects. Left: flow velocity profiles; Right: \ac{CIR} as calculated in Eqs. (\ref{eq:cir_newtonian}), (\ref{eq:cir_pl}), and (\ref{eq:cir_hb}).}
    \label{fig:non_newtonian_models}
    \vspace{-0.6cm}
\end{figure}
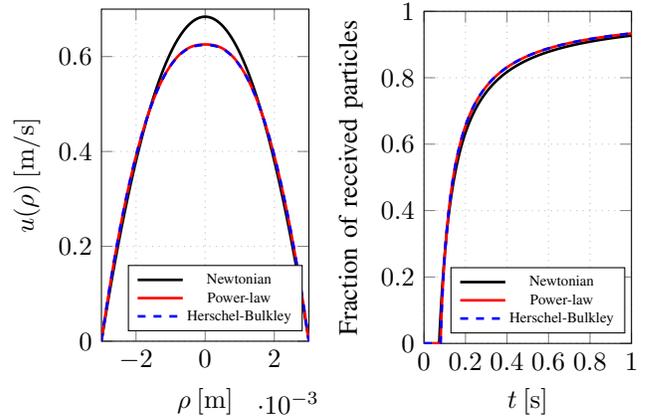
\begin{figure}
    \centering
    \begin{tikzpicture}[scale=0.7, every node/.style={transform shape}]
    
    \begin{axis}[
        xlabel={$\rho \: [\SI{}{\meter}]$},
        ylabel={$u(\rho) \: [\SI{}{\meter\per\second}]$},
        legend pos=north west,
        legend style={nodes={scale=0.8, transform shape}},
        width=\columnwidth,
        grid style=dotted,
        grid=both,
        yticklabel style={/pgf/number format/fixed},
        xticklabel style={/pgf/number format/fixed}, 
        axis lines=box,
        xmin = -3e-3,
        xmax = 3e-3,
        ymin = 0,
        ymax = 1.0,
    ]
    
    \addplot [mark=none, color=black, line width = 1.5pt] table [x expr=\thisrowno{0}, y expr=\thisrowno{1}, col sep=comma] {figures/profile_001.csv};
    \addlegendentry{Sim. $x = \SI{0.01}{\meter}$}

    \addplot [mark=none, color=black, line width = 1.5pt, dashed] table [x expr=\thisrowno{0}, y expr=\thisrowno{1}, col sep=comma] {figures/profile_0025.csv};
    \addlegendentry{Sim. $x = \SI{0.025}{\meter}$}

    \addplot [mark=none, color=black, line width = 1.5pt, dotted] table [x expr=\thisrowno{0}, y expr=\thisrowno{1}, col sep=comma] {figures/profile_004.csv};
    \addlegendentry{Sim. $x = \SI{0.04}{\meter}$}

    \addplot [mark=none, color=red, line width = 1.5pt] table [x expr=\thisrowno{0}, y expr=\thisrowno{1}, col sep=comma] {figures/profile_power_law.csv};
    \addlegendentry{Power-law}

    \addplot [mark=none, color=blue, line width = 1.5pt, dashed] table [x expr=\thisrowno{0}, y expr=\thisrowno{1}, col sep=comma] {figures/profile_herschel_bulkley.csv};
    \addlegendentry{Herschel-Bulkley}
    
    \end{axis}
    \end{tikzpicture}
    \vspace{-0.2cm}
    \caption{Comparison of different flow velocity profiles obtained from the OpenFOAM simulation at three points in the channel (in front of, at, and behind the plaque). The two non-Newtonian analytical models have been fitted to the case for $x=\SI{0.01}{\meter}$. }
    \label{fig:profile_comparison}
    \vspace{-0.4cm}
\end{figure}
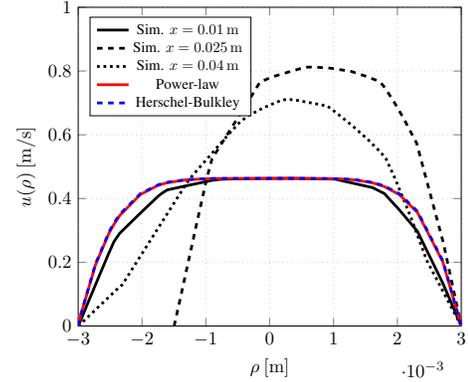

We will now look at the results from our derivations and compare the models based on flow profile and \ac{CIR}. The following parameter values are assumed: Physical parameters from Section~\ref{sec:system_model}, and fluid model parameters $n = 0.708$~\cite{shibeshiRheologyBloodFlow2005}, $\tau_\mathrm{y} = \SI{5e-3}{\pascal}$~\cite{merrillRheologyHumanBlood1963}, $K = \SI{17e-3}{\pascal^n}$~\cite{shibeshiRheologyBloodFlow2005}. In Fig.~\ref{fig:non_newtonian_models}, the resulting flow profiles for the three models are shown on the left side, and on the right side, the corresponding \acp{CIR}. We can clearly see a difference between the Newtonian and the two non-Newtonian models, particularly for the flow profile. The non-Newtonian models form the characteristic flattened plug flow profile, as mentioned at the beginning of the section, caused by the presence of smaller and much larger particles in blood~\cite{ashrafmansouriMathematicalModelingMicro2024}. The effect on the \acp{CIR} is less noticeable in comparison. The non-Newtonian models result in a slightly later arrival of the earliest particles. Still, the higher velocity towards the sides causes the fraction of received particles to rise more quickly in the middle part of the \ac{CIR}. We observe only very tiny deviations between the power-law and Herschel-Bulkley model, suggesting that in the regime of parameters that our scenario operates, the different non-Newtonian models are practically equivalent, and using any one is valid.
Lastly, in Fig.~\ref{fig:profile_comparison}, we compare the analytical flow profiles to the flow profiles acquired from the simulation of the plaque according to the description in Section~\ref{sec:system_model} and for a plaque of size $r_p = 0.25\times r_c$. The flow profile is shown at three different points, one in front of the plaque ($x=\SI{0.01}{\meter})$, directly at the plaque ($\SI{0.025}{\meter}$), and behind the plaque ($\SI{0.04}{\meter}$). The models are shown using the parameters according to the point in front of the plaque. We observe two essential things. Firstly, while the analytical models roughly capture the shape of the profile correctly, and the simulated profile also exhibits the plug flow phenomenon, there still is a noticeable deviation between the analytical and simulation results, indicating that simulations are necessary to capture all the effects in the considered scenario. Secondly, we can compare the shapes of the flow profiles at the different points in the channel. Looking at the flow profile next to the plaque, the flow speed increases as expected, specifically by a factor of $\frac{\SI{81e-3}{\meter\per\second}}{\SI{46e-3}{\meter\per\second}} = 1.77$, which is almost exactly the predicted increase of the flow speed, due to our Venturi effect model, i.e., $\frac{r_c^2}{r(x)^2} = \frac{1}{0.75^2} = 1.78$. Crucially, the flow profile is altered in a lasting way, as it is significantly skewed and asymmetric even after passing the plaque. This reinforces the fact that only simulation can capture the scenario accurately, and analytical models are only able to make rough predictions. Secondly, this represents another possible metric for plaque detection, as uneven, skewed flow patterns might be detectable over more extended time periods by \acp{BNM} that move and communicate around the plaque. This result also tracks with in-vivo measurements that have previously found skewed asymmetric flow profiles in the presence of plaque and more smooth and consistent results after plaque removal~\cite{kamenskiyVivoThreedimensionalBlood2011}.

\section{Pulsatile Inlet Velocity Simulation Approach}

To extend our \ac{CFD} simulations, we applied a human carotid artery pulsatile flow profile as an inlet velocity boundary condition on top of our previous work in~\cite{Hofmann2024} to be able to map future \ac{IoBNT} applications more realistically.
Since the time the particles are released from the circular \ac{TX} in the pulsatile profile significantly influences the \ac{CIR}, we consider three different time points for releasing the particles in our analysis, cf. Fig.~\ref{fig:pulsatile}.
Following~\cite{kamenskiyVivoThreedimensionalBlood2011}, the particles are released at the corresponding time instances for the \ac{PS}, the \ac{ED}, and the \ac{LD}, marked as $t_{\mathrm{PS}} = \SI{0.16}{\second}$, $t_{\mathrm{ED}} = \SI{0.4}{\second}$, and $t_{\mathrm{LD}} = \SI{0.9}{\second}$ as well as dashed red lines in Fig.~\ref{fig:pulsatile}.
% The times for releasing the particles on the \ac{TX} side are as follows: $t_{\mathrm{PS}} = \SI{0.16}{\second}$, $t_{\mathrm{ED}} = \SI{0.4}{\second}$, and $t_{\mathrm{LD}} = \SI{0.9}{\second}$.

%Following~\cite{Hofmann2024}, we assume \acp{SPION} as information carrier particles. 

\subsection{Simulation Control}

As proposed in our previous work~\cite{Hofmann2024OpenFOAM}, we utilize the \ac{MPPICFoam} solver in the OpenFOAM software suite for the simulations.
%Zhou \emph{et al.}~\cite{Zhou2024OpenFOAM} discussed the suitability of the \ac{MPPICFoam} solver in the flow-dominated and the dispersion regime, stating that the non-modified \ac{MPPICFoam} solver is only suitable for the flow-dominated regime.
% Following~\cite{jamali2019channel}, the dispersion factor $\alpha$ is given by:

% \begin{equation}
%     \alpha = \frac{Dl_c}{\overline{v}r_c^2},
% \end{equation}

% whereby $D $ denotes the diffusion coefficient of \acp{SPION}, $l_c = \SI{50}{\milli\meter}$, and $r_c = \SI{3}{\milli\meter}$.
% $\overline{v} = \SI{0.0515}{\meter\per\second} $ denotes the weighted averaged velocity of the human carotid pulsatory velocity profile, cf.~Fig~\ref{fig:pulsatile} timely averaged over one cycle.

% The diffusion coefficient of \acp{SPION} can be estimated from their average radius $r_\mathrm{SPION} = \SI{50}{\nano\meter}$~\cite{Bartunik2023channel} and the formula for the diffusion coefficient for particles suspended in a liquid~\cite{jamali2019channel} given by
% \begin{equation}
%     D = \frac{k_\mathrm{B}T}{6\pi\eta r_\mathrm{SPION}} = \SI{1.1e-12}{\square\meter\per\second},
% \end{equation}
% where $k_\mathrm{B} = \SI{1.38e-23}{\joule\per\kelvin}$, $T=\SI{300}{\kelvin}$, and the dynamic viscosity of blood is set to an average value of $\eta = \SI{4e-3}{\pascal\second}$. In this case, the dispersion factor can be calculated to be $\alpha = 2.4\times10^{-7} \ll 1$, signifying the clear validity of the flow-dominated regime~\cite{jamali2019channel}.
We simulate a total time of $t_{\mathrm{release}} = t_{\mathrm{SOI}} + \SI{1}{\second}$, denoting the time after the release of the particles.
\ac{SOI} denotes the release of the particles, i.e., $t_{\mathrm{SOI}} = \{ t_{\mathrm{PS}}, t_{\mathrm{ED}}, t_{\mathrm{LD}} \}.$
The simulation time step size $\Delta t$ is constant, i.e., $\Delta t = \SI{1e-4}{\second}$ for the $r_p = 0 \times r_c$, $r_p = 0.25 \times r_c$, and $r_p = 0.5 \times r_c$ scenario.
Due to the high resolution of the mesh at the narrow point of the plaque for $r_p = 0.75 \times r_c$ and the high \ac{PS} amplitude, here we set $\Delta t = \SI{1e-5}{\second}$.

\subsection{Simulation Results}

%%%%%
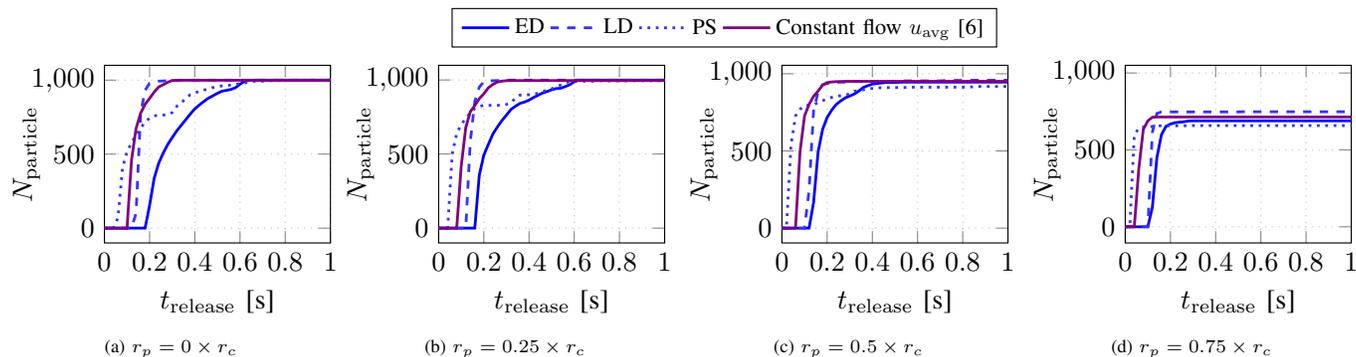
\begin{figure*}[t]
    \centering
    
    % First subplot
    \begin{subfigure}{0.245\textwidth} 
        \centering
        \begin{tikzpicture}[scale=1.05, every node/.style={transform shape}]
            \begin{axis}[
                xlabel={$t_{\mathrm{release}}$ [\SI{}{\second}]},
                ylabel={$N_{\mathrm{particle}}$},
                ylabel shift = -10pt,
                legend pos=south east,
                legend style={nodes={scale=0.8, transform shape}, legend columns=-1, at={(3.95,1.1)}},
                width=\textwidth,
                grid style=dotted,
                grid=both,
                yticklabel style={/pgf/number format/fixed},
                xticklabel style={/pgf/number format/fixed},
                axis lines=box,
                xmin=0.0,
                xmax=1.0,
            ]
            
            \addplot [mark=none, color=blue, line width=1.0pt] table [x=time_new, y=number_of_received_particles_04, col sep=comma] {figures/Plaque_laminar_1_varying.csv};
            \addlegendentry{ED}
            
            \addplot [mark=none, color=blue!80, dashed, line width=1.0pt] table [x=time_new, y=number_of_received_particles_09, col sep=comma] {figures/Plaque_laminar_1_varying.csv};
            \addlegendentry{LD}
            
            \addplot [mark=none, color=blue!80, dotted, line width=1.0pt] table [x=time_new, y=number_of_received_particles_016, col sep=comma] {figures/Plaque_laminar_1_varying.csv};
            \addlegendentry{PS}
            
            \addplot [mark=none, color=violet, line width=1.0pt] table [x=time_new, y=number_of_received_particles_constant, col sep=comma] {figures/Plaque_laminar_1_varying.csv};
            \addlegendentry{Constant flow $u_{\mathrm{avg}}$~\cite{Hofmann2024}}
            
            \end{axis}
        \end{tikzpicture}
        \caption{$r_p = 0 \times r_c$} % Subcaption for the first subplot
        \label{fig:sub1}
    \end{subfigure}%
    %\hspace{0.5cm} % Adjust horizontal space between subfigures if needed
    % Second subplot
    \begin{subfigure}{0.245\textwidth}
        \centering
        \begin{tikzpicture}[scale=1.05, every node/.style={transform shape}]
            \begin{axis}[
                xlabel={$t_{\mathrm{release}}$ [\SI{}{\second}]},
                ylabel={$N_{\mathrm{particle}}$},
                ylabel shift = -10pt,
                legend pos=south east,
                legend style={nodes={scale=0.8, transform shape}},
                width=\textwidth,
                grid style=dotted,
                grid=both,
                yticklabel style={/pgf/number format/fixed},
                xticklabel style={/pgf/number format/fixed},
                axis lines=box,
                xmin=0.0,
                xmax=1.0,
            ]
            
            \addplot [mark=none, color=blue, line width=1.0pt] table [x=time_new, y=number_of_received_particles_04, col sep=comma] {figures/Plaque_laminar_4_varying.csv};
            %\addlegendentry{ED}
            
            \addplot [mark=none, color=blue!80, dashed, line width=1.0pt] table [x=time_new, y=number_of_received_particles_09, col sep=comma] {figures/Plaque_laminar_4_varying.csv};
            %\addlegendentry{LD}
            
            \addplot [mark=none, color=blue!80, dotted, line width=1.0pt] table [x=time_new, y=number_of_received_particles_016, col sep=comma] {figures/Plaque_laminar_4_varying.csv};
            %\addlegendentry{PS}
            
            \addplot [mark=none, color=violet, line width=1.0pt] table [x=time_new, y=number_of_received_particles_constant, col sep=comma] {figures/Plaque_laminar_4_varying.csv};
            %\addlegendentry{Constant flow $v_{\mathrm{eff}}$~\cite{Hofmann2024}}
            
            \end{axis}
        \end{tikzpicture}
        \caption{$r_p = 0.25 \times r_c$} 
        \label{fig:sub2}
    \end{subfigure}
\begin{subfigure}{0.245\textwidth}
        \centering
        \begin{tikzpicture}[scale=1.05, every node/.style={transform shape}]
            \begin{axis}[
                xlabel={$t_{\mathrm{release}}$ [\SI{}{\second}]},
                ylabel={$N_{\mathrm{particle}}$},
                ylabel shift = -10pt,
                legend pos=south east,
                legend style={nodes={scale=0.8, transform shape}},
                width=\textwidth,
                grid style=dotted,
                grid=both,
                yticklabel style={/pgf/number format/fixed},
                xticklabel style={/pgf/number format/fixed},
                axis lines=box,
                xmin=0.0,
                xmax=1.0,
            ]
            
            \addplot [mark=none, color=blue, line width=1.0pt] table [x=time_new, y=number_of_received_particles_04, col sep=comma] {figures/Plaque_laminar_7_varying.csv};
            %\addlegendentry{ED}
            
            \addplot [mark=none, color=blue!80, dashed, line width=1.0pt] table [x=time_new, y=number_of_received_particles_09, col sep=comma] {figures/Plaque_laminar_7_varying.csv};
            %\addlegendentry{LD}
            
            \addplot [mark=none, color=blue!80, dotted, line width=1.0pt] table [x=time_new, y=number_of_received_particles_016, col sep=comma] {figures/Plaque_laminar_7_varying.csv};
            %\addlegendentry{PS}
            
            \addplot [mark=none, color=violet, line width=1.0pt] table [x=time_new, y=number_of_received_particles_constant, col sep=comma] {figures/Plaque_laminar_7_varying.csv};
            %\addlegendentry{Constant flow $v_{\mathrm{eff}}$~\cite{Hofmann2024}}
            
            \end{axis}
        \end{tikzpicture}
        \caption{$r_p = 0.5 \times r_c$} 
        \label{fig:sub3}
    \end{subfigure}
    \begin{subfigure}{0.245\textwidth}
        \centering
        \begin{tikzpicture}[scale=1.05, every node/.style={transform shape}]
            \begin{axis}[
                xlabel={$t_{\mathrm{release}}$ [\SI{}{\second}]},
                ylabel={$N_{\mathrm{particle}}$},
                ylabel shift = -10pt,
                legend pos=south east,
                legend style={nodes={scale=0.8, transform shape}},
                width=\textwidth,
                grid style=dotted,
                grid=both,
                yticklabel style={/pgf/number format/fixed},
                xticklabel style={/pgf/number format/fixed},
                axis lines=box,
                xmin=0.0,
                xmax=1.0,
                ymax = 1050,
            ]
            
             \addplot [mark=none, color=blue, line width=1.0pt] table [x=time_new, y=number_of_received_particles_04, col sep=comma] {figures/Plaque_laminar_10_varying.csv};
            % \addlegendentry{ED}
            
             \addplot [mark=none, color=blue!80, dashed, line width=1.0pt] table [x=time_new, y=number_of_received_particles_09, col sep=comma] {figures/Plaque_laminar_10_varying.csv};
            % \addlegendentry{LD}
            
             \addplot [mark=none, color=blue!80, dotted, line width=1.0pt] table [x=time_new, y=number_of_received_particles_016, col sep=comma] {figures/Plaque_laminar_10_varying.csv};
            % \addlegendentry{PS}
            
             \addplot [mark=none, color=violet, line width=1.0pt] table [x=time_new, y=number_of_received_particles_constant, col sep=comma] {figures/Plaque_laminar_10_varying.csv};
            % \addlegendentry{Constant flow $v_{\mathrm{eff}}$~\cite{Hofmann2024}}
            
            \end{axis}
        \end{tikzpicture}
        \caption{$r_p = 0.75 \times r_c$} 
        \label{fig:sub4}
    \end{subfigure}
    \vspace{-0.2cm}
    \caption{Total number of received particles $N_{\mathrm{particle}}$ for different radial extensions of the plaque $r_p$ for different points of release in time, i.e., $t_{\mathrm{ED}}$, $t_{\mathrm{LD}}$, and $t_{\mathrm{PS}}$. $t_{\mathrm{release}}$ denotes the time after the release of the molecules at the planar \ac{TX}. A total number of 1000 particles is released by the circular \ac{TX}.}
    \label{fig:CIR}
    \vspace{-0.4cm}
\end{figure*}

%%%%%

To evaluate the performance of the simulated scenarios, i.e., different release times of the particles and different plaque expansions, we plot the number of the received particles $N_{\mathrm{particles}}$ over the time after the release, see Fig.~\ref{fig:CIR}.
For benchmarking, $N_{\mathrm{particles}}$ is also plotted for the default scenario from our previous work in~\cite{Hofmann2024}, denoted as 'Constant flow $u_{\mathrm{avg}}$' (purple solid line).

As observed in~\cite{Hofmann2024}, Fig.~\ref{fig:CIR} demonstrates that with increasing expansion of the plaque in the radial direction, i.e., with a more substantial influence of the Venturi effect, the first particles are more likely to reach the \ac{RX} on average, even with applied pulsatile inlet velocity boundary condition. Specifically, the traversal speed increases by 5.5\% for $r_p = 0.25\times r_c$, 33\% for $r_p = 0.5\times r_c$, and 43\% for $r_p = 0.75\times r_c$. Comparing this with our analytical Venturi model in Eq. (\ref{eq:venturi}), the predicted values are 15\%, 30\%, and 43\%, respectively, which align quite well with the simulation, especially for larger plaques, for which the more abrupt transition might cause the piecewise velocity assumption of the model to be more valid.

The subplots (a) to (d) in Fig.~\ref{fig:CIR} further demonstrate that the first particles released at $t_{\mathrm{PS}}$ arrive faster at the \ac{RX}, due to the high amplitude at $t_{\mathrm{PS}}$, cf. Fig.~\ref{fig:pulsatile}, enabling particles to overcome the obstructions caused by plaque formation more efficiently. 
However, due to the steep drop in the amplitude after the global maximum, the last particles arrive much later than the particles released at $t_{\mathrm{ED}}$ and $t_{\mathrm{LD}}$. 
This pulsatile inlet flow condition behavior can also be seen in the irregularity in the upper third of the \ac{PS} plot curve in Fig.~\ref{fig:CIR}. 
In all plots in Fig.~\ref{fig:CIR}, the number of received particles for the release time $t_{\mathrm{LD}}$ approaches the default scenario, initially with a lower number of received particles, but with increasing time $t_{\mathrm{release}}$ with a higher number of received particles. 
This observed behavior can also be explained by the pulsatile inlet flow velocity profile, as the amplitude of the flow rate at the release time is below the constant amplitude of $u_{\mathrm{avg}}$ but increases sharply after approximately \SI{0.1}{\second} due to the systolic behavior.
Finally, for the release time $t_{\mathrm{ED}}$, Fig.~\ref{fig:CIR} demonstrates that due to the low amplitude of the beginning diastole, the particles arrive later at the \ac{TX} compared to the other simulated scenarios. 

The default scenario~\cite{Hofmann2024} represents the scenario without pulsatile effects, where particles experience a steady flow. 
The corresponding curve remains relatively consistent across all plaque expansions and generally exhibits slower initial reception than \ac{PS} scenario but faster than \ac{ED} and \ac{LD} scenarios.
In cases with large plaque ($r_p = 0.75 \times r_c$), the default scenario shows a slower increase in the number of received particles compared to the \ac{PS} scenario, suggesting that the systolic phase in a pulsatile flow offers a higher particle transport efficiency even in obstructed channels.

Overall, the \ac{PS} scenario demonstrates the highest transmission speed in overcoming plaque-induced obstacles.
In contrast, \ac{ED} releasing particles results in significant delays, emphasizing the importance of synchronizing the particle release with pulsatile phases in \ac{MC} \ac{IoBNT} applications.
In general, in the \ac{PS} scenario, we also see the largest difference in the number of received particles, as well as the shape of the \ac{CIR}, implying increased suitability for detecting the plaque.

\section{Conclusion and Outlook}

In this paper, we analyzed a human carotid blood vessel and plaque obstruction model, focusing on the non-Newtonian behavior of blood, modeling its flow as an \ac{MC} channel to derive \acp{CIR} at the \ac{RX}.
Furthermore, based on our previous work in~\cite{Hofmann2024}, we modeled plaque formations using a \ac{CFD} tool, implementing a pulsatile inlet velocity flow profile mimicking a human carotid artery. 
The simulation results demonstrate how the timing of the particle release within the cardiac cycle and the presence of plaque formations affect the \ac{MC} system performance, evaluated by the \ac{CIR}. 

Future research includes turbulence modeling of the envisioned scenarios as the plaque formation can lead to turbulence, especially under pulsatile inlet flow boundary conditions. 
Furthermore, \ac{FSI} is to be considered and implemented for future research. 
%To the best of the authors' knowledge, there are no publications in \ac{MC} and \ac{FSI}. 
A first trial using the OpenFOAM toolbox solids4foam~\cite{Cardiff2018} can be found in the published dataset~\cite{Hofmann2024dataet}. 
The considered \ac{FSI} scenario includes the obstacle-free channel from Fig.~\ref{fig:plaque_schematic}, with a pulsatile inlet flow boundary condition.  

% % use section* for acknowledgment
% \section*{Acknowledgment}

% This work was supported by the German Research Foundation (DFG) as part of Germany's Excellence Strategy—EXC 2050/1—Cluster of Excellence "Centre for Tactile Internet with Human-in-the-Loop" (CeTI) of Technische Universität Dresden under project ID 390696704 and the Federal Ministry of Education and Research (BMBF) in the programme of "Souverän. Digital. Vernetzt." Joint project 6G-life, grant numbers 16KISK001K and 16KISK002. 
% This work was also partly supported by the project IoBNT, funded by the German Federal Ministry of Education and Research (BMBF) under grant number 16KIS1994.

\bibliographystyle{ieeetr}
\bibliography{references}

\end{document}